\documentstyle[aps,epsfig]{revtex}

\begin{document}

\title{Bose-Einstein Condensation on inhomogeneous networks: mesoscopic aspects
versus thermodynamic limit}
\author{P. Buonsante \footnote{buonsant@fis.unipr.it}$^{1,2,3}$,
R. Burioni$^{1,2}$  \footnote{burioni@fis.unipr.it}, D.
Cassi$^{1,2}$ \footnote{cassi@fis.unipr.it} and A. Vezzani$^{1,2}$
\footnote{vezzani@fis.unipr.it}}

\address{ $^1$ Istituto Nazionale Fisica della Materia (INFM)\\
$^2$ Dipartimento di Fisica, Universit\`a di Parma, parco Area delle Scienze 7A
43100 Parma Italy\\  $^3$ Dipartimento di Fisica, Politecnico di
Torino, corso Duca degli Abruzzi 24 10129 Torino Italy}

\maketitle

\begin{abstract}

We study the filling of states in a  pure hopping boson model on the
comb lattice, a low dimensional discrete structure where geometrical
inhomogeneity induces 
Bose-Einstein condensation (BEC) at finite temperature. By a careful
analysis of the thermodynamic limit on combs we show that, unlike the
standard lattice case, BEC is characterized by a macroscopic occupation of
a finite number of states with energy belonging to a small neighborhood
of the ground state energy. Such remarkable
feature gives rise to an anomalous behaviour in the large distance two-point
correlation functions. Finally, we prove a general theorem providing
the conditions for the pure
hopping model to exhibit the standard behaviour, i.e. to present a macroscopic
occupation of the ground state only.

\end{abstract}
\maketitle
\section{Introduction}

The most recent experiments on Bose-Einstein condensation (BEC) stimulated a
large wealth of theoretical work aimed at a better understanding of the basic
properties of such interesting phenomenon \cite{bec}. In
particular the possibility of confining ultra-cold bosonic atoms within
optical lattices, along with the most recent studies on arrays of Josephson
junctions, arouse a great interest in bosonic models defined on Euclidean
lattices \cite{optlat,jja}.

Even more interesting from this point of view would be the arrangement
of the Josephson
junctions, or possibly optical traps, into more complex networks \cite{exp}.
Indeed it was recently put into evidence
\cite{epl,jpb,jpa} that, due to topological inhomogeneity, BEC at finite
temperature can occur on low dimensional structures, such as the comb lattice
\cite{pet}, even in
the absence of external potential. Such results were obtained mainly in the
thermodynamic limit (i.e. for structures of infinite size), but it is clear
that a deeper understanding of the onset of the phenomenon on finite-size
structures is needed. This is especially true in view of the experimental
research which is currently being developed in this field \cite{exp}.

One of the distinctive features of usual BEC
(in continuous Euclidean geometry,
possibly in the presence of harmonic potentials) is the macroscopic occupation
of a single quantum state. More precisely it is possible to prove that the
filling of any excited state vanishes in the thermodynamic limit.
As for general networks, Refs. \cite{epl,jpb,jpa} mainly deal with
the thermodynamic aspect of the
problem. In particular it is shown that on a low dimensional inhomogeneous
network, such as the comb lattice, BEC is characterized by the macroscopic
occupation of the states belonging to an arbitrarily small energy neighborhood
of the ground state. This is a more general condition, since it does not
necessarily entail that only the ground state features a macroscopic
occupation.

In the following we indeed show that the macroscopic
occupation of the
ground state does not entirely account for BEC on the comb lattice.
Even in the thermodynamic limit, the filling of the structure is completely
described only if a finite number of states belonging to a small neighborhood
of the ground state is considered.
This result is proven by exactly evaluating the occupation of the
quantum states of mesoscopic comb lattices, with large but finite size.
The spectrum of such structures is
nearly continuous, but the energy levels are still distinguishable since they
pertain to orthogonal wave functions which differ on a macroscopic scale.

The paper is organized as follows. In Section II we introduce the
pure hopping model \cite{epl} which describes the low coupling
limit for the Josephson junction arrays or a model of non
interacting atoms confined in an optical lattice. Using the Van
Hove spheres, we define the thermodynamic limit for a general infinite
discrete structure. Then we recall the definition of {\it the low energy
hidden spectrum} and we give the general conditions on the
graph spectra for the condensation in an 
arbitrarily small energy region \cite{jpa}. 
In Section III we consider the problem of the
states filling for the pure hopping bosonic model comparing the
known results for the 3-dimensional lattice with the numerical
results obtained on the comb graph. We show that in the first case
only the lowest energy state presents a macroscopic occupation,
while the comb graph exhibits a different behaviour since a
macroscopic occupation of many states is present. In Section IV we
solve analytically the problem of the states filling on the comb
graph proving that, in the thermodynamic limit, also the first
excited states are filled by a finite fraction of particles.
Furthermore we show that the numerical data perfectly fit the
analytical predictions. In Section V we prove that the macroscopic
filling of many states gives rise to an anomalous behaviour of the
large distance two point correlation functions. In 
Section VI we give the spectral conditions for the pure hopping model
to present a macroscopical occupation of a single quantum state and in
Section VII we present our conclusions.

\section{The bosonic hopping models on graphs}

Graph theory provides the most natural mathematical description
of generic discrete networks \cite{graph}. A graph is
a countable set $V$ of sites $i$ connected
pairwise by a set $E$ of unoriented links $(i,j)=(j,i)$. Two sites
joined by a link are called nearest neighbours.  The topology of a graph is
fully described by its adjacency matrix $A_{ij}$, with $A_{ij}=1$ if $(i,j)$ is
a link of the graph and $A_{ij}=0$ otherwise.
A {\it walk} in  ${G}$ is a sequence of concatenated
links $\{(i,k),(k,h)\ldots,(n,m),(m,k)\}$ and a graph is
said to be connected if for any two sites there is always a walk joining them.
The length of a walk is the number of links appearing in the
sequence, and the length of the shortest walk joining two sites is
called the {\it chemical distance} between them. The latter defines the
intrinsic metric every connected graph is endowed with. The Van Hove sphere
$S_{r,o}$  of center $o$ and radius $r$ is the set of  sites
whose chemical distance from site $o$ is equal or less than $r$.
In the following we will consider only connected graphs with polynomial growth,
where the number
$N_{r,o}$ of sites within $S_{r,o}$  grows at most as a power of the radius.
This choice ensures that the discrete structure can be embedded in a finite
dimensional Euclidean space. Since for this class of graphs
it is possible to prove that the thermodynamic limit is independent from the
choice of the center of the sphere \cite{rimtim}, in the following we will
drop the relevant subscript.

Let us recall some basic results obtained for the pure hopping
model on a generic graph (a complete description of the subject can be found in
\cite{jpb,jpa}).
On a graph $G$ the pure hopping model for bosons is defined by the Hamiltonian
\cite{epl}:
\begin{equation}
{{H}}=-t\sum_{i,j\in V} A_{ij}a^{\dag}_ia_j
\label{ham}
\end{equation}
where $a^{\dag}_i$ and $a_i$ are the creation and annihilation operator
at site $i$ ($[a_i,a^{\dag}_j]=\delta_{ij}$) and $A_{ij}$ is the adjacency
matrix of the graph.

To study the behaviour of (\ref{ham}) in the thermodynamic limit,
where BEC can arise, we restrict the model to $S_{r}$ and analyze its
properties as the radius of  the Van Hove Sphere goes to infinity.
The model restricted to $S_r$ is defined by the Hamiltonian
\begin{equation}
H^{r}=\sum_{i,j} -tA^r_{ij}a^{\dag}_ia_j
\label{hamsr}
\end{equation}
where $A^r_{ij}=A_{ij}$ if $i,j\in S_r$ and
$A^{r}_{ij}=0$ otherwise. The corresponding normalized density of states
$\rho^r(E)$ is
\begin{equation}
\rho^r(E)={1\over{N_r}}\sum_k \delta(E-E^r_k)
\label{rhor}
\end{equation}
where $E^r_k$ are the eigenvalues of $-tA^r_{ij}$ and $N_r$ is the number of
sites within the sphere. The function $\rho(E)$ is defined to be the
thermodynamic density of states of $-tA_{ij}$ if it satisfies the following
condition:
\begin{equation}
\lim_{r\to\infty} \int |\rho^r(E)-\rho(E)| dE =0.
\label{dens}
\end{equation}
In general the asymptotic behavior at low energies of the thermodynamic density
of states is described by a power law of the form
\begin{equation}
\label{asbeh}
\rho(E) \sim (E-E_m)^{{\alpha\over 2}-1} \ \ \ \ \ \ {\rm for} \ E\to E_m
\label{defd}
\end{equation}
where $E_m \equiv {\rm Inf} ({\rm Supp}(\rho(E)))$ and $\alpha$ is an exponent
determined by the topology of the graph \cite{ao,univ}.

On inhomogeneous structures the density of states can present some interesting
anomalies, such as the {\it hidden region} of the  spectrum defined in \cite{jpb}
and \cite{jpa}. A {\it hidden region} of the  spectrum consists of an energy
interval $[E_1,E_2]$ such that $[E_1,E_2]\cap {\rm Supp}(\rho(E))= \emptyset$
and $\lim_{r\to \infty} N^r_{[E_1,E_2]}>0$, where  $N^r_{[E_1,E_2]}$ is the
number of eigenvalues of $-tA^r_{ij}$ in the interval $[E_1,E_2]$. Notice that
in general $N^r_{[E_1,E_2]}$ can diverge for $r\to \infty$ and the eigenvalues
can become dense in $[E_1,E_2]$ in the thermodynamic limit. Therefore the 
presence of a hidden spectrum is a far more general property
than the existence of a discrete spectrum, where a finite
number of states fills the spectral region.
An interesting example of this new kind of behaviour is exhibited by the the
comb lattice \cite{epl,jpb} (see figure 1) which will be studied in
detail in the following sections. We now define the lowest energy level for the
sequence of densities $\rho_r(E)$, setting $E_0^r={\rm Inf}_k (E_k^r)$ and
$E_0=\lim_{r \to \infty} E_0^r$. In general, $E_0\leq E_m$.
If $E_0<E_m$, then $[E_0,E_m]$ is a hidden region of the spectrum
which will be called low energy hidden spectrum.
\begin{figure}
\begin{center}
\epsfig{file=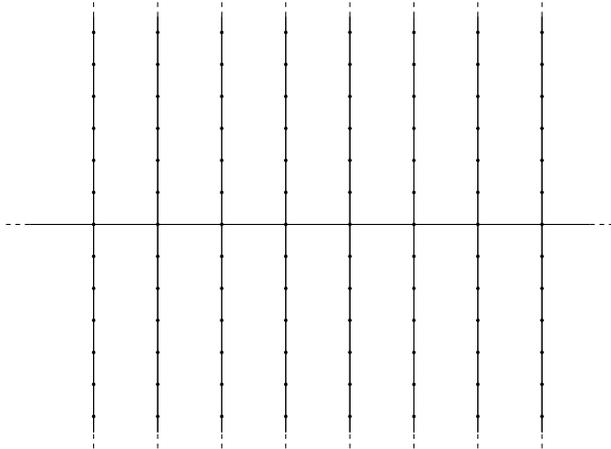, height=6cm,angle=0}
\label{fcomb}
\caption{The comb graph, which is obtained connecting to each  
site of a linear chain, called backbone, a 1-dimensional chain called  
finger.}
\end{center}
\end{figure}
The correct way of taking the thermodynamic limit consists in adjusting the
population $N$ of the system so that the filling $N/N_r$ is set to a fixed
value $f$ as the radius of the Van Hove sphere goes to infinity.
In particular, in the macro-canonical ensemble, the equation that determines
the fugacity $z$
as a function of $\beta=T^{-1}$, $f$ and $r$ is:
\begin{equation}
f= \int { 1\over z^{-1}e^{\beta E}-1}\rho^r(E) dE.
\label{eq1}
\end{equation}
Setting $E_0=0$ yields $0\leq z(f,\beta,r) \leq 1$.
A system presents BEC at finite temperature if there exists a
temperature $T_C$ such that for all $T\leq T_C$  $\lim_{r\to
\infty}z(f,\beta,r)=1$.

The general conditions for the occurrence of BEC at finite temperature is
strictly related to the properties of $\rho(E)$  \cite{jpa}. In particular it is
proven that BEC arises either in models presenting a low energy hidden spectrum
or in models where the parameter $\alpha$ appearing in equation (\ref{asbeh}) is
larger than two. The comb lattice is an example of the former situation,
whereas the latter condition is satisfied by any Euclidean lattice with
dimension larger than two, since for such a structure
the parameter $\alpha$ coincides with the Euclidean dimension.

\section{The single state filling}

BEC on Euclidean lattices is characterized by the non-analytic behaviour
of the thermodynamic functions at the critical temperature and by a
macroscopic filling of a single quantum state.
The definition of BEC we gave in the previous section is strictly connected with
the singular properties at $T_C$. For example in the thermodynamic limit
the average energy per particle is:
\begin{equation}
\langle E\rangle= {1\over f}\int {E\over z^{-1}e^{\beta E}-1}\rho(E) dE
\label{energy}
\end{equation}
where $z=1$ for temperature below the critical temperature, whereas  for $T>T_C$
it is a function of $T$ and $f$ determined by (\ref{eq1}). Therefore the
equation (\ref{energy}) presents the typical non-analytic behaviour of the
thermodynamic functions of BEC at the critical point.

Let us now consider the filling of the states. Below $T_C$ any arbitrarily
small energy region around $E_0$ is filled by a finite fraction of particles,
as it is known by classical results on BEC.
More precisely, for all $\epsilon>0$ the fraction $n_{\epsilon}$ of particles
with  energy in the region $[E_0,E_0+\epsilon]$ is always greater than the
positive quantity $n_{TD}$:
\begin{equation}
n_{TD}=f- \int { 1\over e^{\beta E}-1}\rho(E) dE
\label{nTD}
\end{equation}
where $E_0=0$ and the subscript $TD$ stands for thermodynamic.

Since $E_0$ is the only real number belonging to $[E_0,E_0+\epsilon]$ for any
value of $\epsilon$, one would argue that the only possible way to satisfy
the condition $n_{\epsilon}>n_{TD}>0$ $\forall \epsilon>0$ is to fill the lowest
energy state with a finite fraction of particles. However if we look carefully
to how the thermodynamic limit is performed, the question is not so trivial.
Indeed, if the model does not present a gap (this is also the case of the usual
condensation on lattices), the energy of the first excited states tends 
to $E_0$
in the thermodynamic limit and it belongs to $[E_0,E_0+\epsilon]$ for all
values of $\epsilon$.
For instance, in order to determine the filling $n_1$ of the first excited state
$E_1$, one first has to solve equation (\ref{eq1}) for $z(r)$ and then evaluate $n_1=\lim_{r\to\infty}N_r^{-1}(z(r)^{-1}\exp(\beta E_1(r))-1)^{-1}$.
The result of this limit depends on how $N_r\to\infty$, $E_1(r)\to E_0=0$ and
$z(r)\to 1$ in the thermodynamic limit.

This limit is trivial on lattices: there the pure hopping model does not present
a gap and it is known that below $T_C$ the lowest energy eigenstate is the only
one with a macroscopic occupation \cite{path}.
Let us first consider the case of the three-dimensional lattice. The
spectrum of a finite lattice of $L^3=N_r$ sites is given by:
\begin{equation}
\sigma_l=\left\{6t -
2t\,\cos \frac{2 \pi k}{L}-
2t\,\cos \frac{2 \pi h}{L}-2t\,\cos \frac{2 \pi j}{L}
\right\}_{k=1,\dots,L;h=1,\dots,L;j=1,\dots,L}
\label{spec3d}
\end{equation}
the equation determining the fugacity $z$ for each finite lattice is:
\begin{equation}
f=\frac{1}{L^3}{z\over 1-z}+
\frac{1}{L^3} \sum_{E_k \in \sigma_l}^{E_k\neq 0} \frac{z}{e^{\beta E_k}-z}
\label{fill3d}
\end{equation}
where $n_0=L^{-3}(z^{-1}-1)^{-1}$ is the filling of the ground state and
$L^{-3}(z^{-1}\exp(\beta E_k)-1)^{-1}$ is the filling of a state of 
energy $E_k$.
Below the critical temperature equation (\ref{fill3d}) has a solution in the
thermodynamic limit only if $z\to 1$ for $L\to\infty$. 
In particular one has that:
\begin{equation}
z \sim 1-\frac{\delta}{L^3} \ \ \ {\rm for\ \ }L\to\infty
\label{zcomb}
\end{equation}
Substituting (\ref{zcomb}) in (\ref{fill3d}) one obtains $n_0=\delta^{-1}$ and
\begin{equation}
n_0=f-\int { 1\over e^{\beta E}-1}\rho(E) dE.
\label{fill3db}
\end{equation}
Hence $n_0=n_{TD}$ and the lowest energy state is the only macroscopically
filled state. As a further check of this result we can explicitly
evaluate  the filling $n_1$ of the first excited state in the thermodynamic
limit. According to equation (\ref{spec3d}) the energy
of the first excited state is $E_1=t(2-2\cos(2\pi/L))$ and, from
equation (\ref{zcomb}),  its filling  is
\begin{equation}
n_1=\lim_{L\to\infty}{1\over L^3}{z\over z^{-1}e^{\beta E_1}-1}=
\lim_{L\to\infty}{1\over L^3}{1\over \beta E_1+\delta/L^3}=0
\label{fill3dc}
\end{equation}
in the thermodynamic limit.

In figure 2 we plotted $n_{TD}$ (obtained from (\ref{nTD}) with
$f=1$), $n_0$  (the filling of the lowest energy state), and $n_{\epsilon}$ for
two different values of  $\epsilon$. To determine $n_0$ and
$n_{\epsilon}$ we first evaluated the exact spectrum of a finite lattice
consisting of $N_r=125000$ sites, then we obtained $z(1,\beta,r)$ by numerical
inversion of equation (\ref{eq1}). This allowed us to evaluate the filling
$n_k=N_r^{-1}(z^{-1}e^{\beta E_k}-1)^{-1}$ of the
energy level $E_k$. From figure 2
we have that the differences between the plots are very small and they can be
ascribed to finite size effects. These numerical results confirm the known
property of BEC on regular lattices of presenting a macroscopic occupation
only in the state of lowest energy.

\begin{figure}
\begin{center}
\epsfig{file=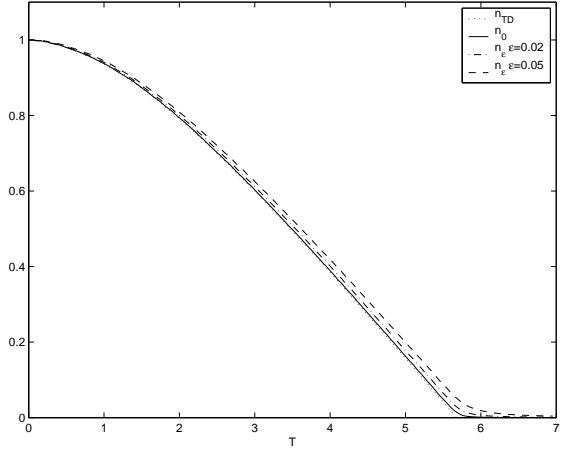, height=6cm,angle=0}
\caption{The state filling in the 3-dimensional lattice as a function of the
temperature measured in units of $t$. The dotted line represents
the theoretical value of $n_{TD}$ obtained from (\ref{nTD}). The solid line is
a numerical evaluation of the filling of the ground state $n_0$. The dashed and
dashed-dotted lines are a numerical evaluation $n_{\epsilon}$ for two different
small values of $\epsilon$: $\epsilon=0.02 t$, $\epsilon=0.05 t$. Numerical data
are obtained from a lattices of $N_r=125000$ sites.}
\end{center}
\end{figure}

Let us focus on the case of the comb graph proving that on
inhomogeneous structures, due to the presence of hidden regions in
the spectrum, a macroscopic filling of states of energy arbitrarily
near to the ground state is possible.
In figure 3 $n_{TD},$ $n_0$ and $n_{\epsilon}$ for a  comb graph
consisting of $40000$ sites are shown. In this case it is evident
that, since the filling of the ground state (the curve $n_0$) is
lower than $n_{TD}$,  there must be macroscopically filled states
other than $E_0$. From the general result
(\ref{nTD}) the energies of these states must tend to $E_0$ as
$L\to\infty$. Actually, the curves $n_{\epsilon}$ representing the
filling relevant to the energy region $[E_0,E_0+\epsilon]$ is very
close to $n_{TD}$ even for small values of $\epsilon$.

\begin{figure}
\begin{center}
\epsfig{file=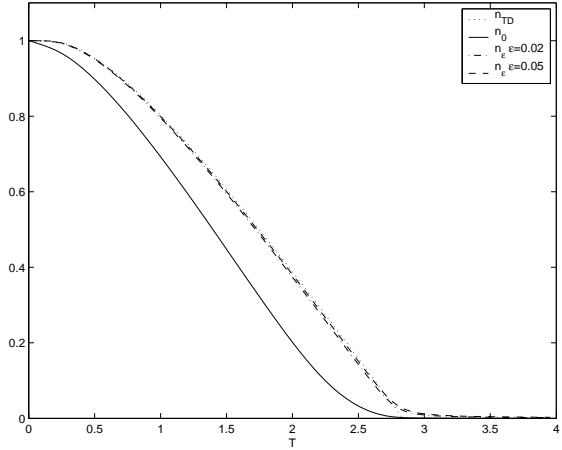, height=6cm,angle=0}
\caption{The state filling in the comb lattice as a function of the
temperature measured in units of $t$. The dotted line represents
the theoretical value of $n_{TD}$ obtained from (\ref{nTD}). The solid line is
a numerical evaluation of the filling of the ground state $n_0$. The dashed and
dashed-dotted lines are a numerical evaluation $n_{\epsilon}$ for two different
small values of $\epsilon$: $\epsilon=0.02 t$, $\epsilon=0.05 t$. Numerical data
are obtained from a comb of $N_r=40000$ sites.}
\end{center}
\end{figure}

\section{States filling on the comb graph}

On the comb graph, the anomalous behaviour of the filling can be explained by
carefully considering equation (\ref{eq1}), which determines the fugacity for
each finite-size comb.
The energy spectrum of the pure hopping model on a finite comb graph consisting
of $N_r=L\times L$ is the union of four sets, $\sigma_{\rm P} =
\left\{E_0 \right\} \cup \sigma_- \cup \sigma_0 \cup \sigma_+$ \cite{jpb}.
Setting $E_0=0$:
\begin{eqnarray}
\sigma_- &=& \left\{ E_k =t\left( \sqrt{8} - 2\,\sqrt{1+\cos^2 \frac{2 \pi
k}{L}}\right) \right\}
_{k=1,2,\frac{L-1}{4}} \nonumber \\
\sigma_0 &=& \left\{E_k =t\left(\sqrt{8} - 2\,\cos \frac{2 \pi k}{L}\right)
\right\}_{k=1,2,L-1}  \\
\sigma_+ &=& \left\{E_k=t\left(\sqrt{8} + 2\,\sqrt{1+\cos^2
\left(\frac{2 \pi k}{L}+\frac{\pi}{2}\right)}\right)\right\}
_{k=1,2, \frac{L-1}{4}} \nonumber
\end{eqnarray}
The degeneration $d(E_k)$ of the states of energy $E_k\in \sigma_-$ and
$E_k\in \sigma_+$ is $2$, while for the states of energy $E_k\in\sigma_0$ we
have $d(E_k)=L$.
Hence, in the thermodynamic limit, $\sigma_-$ and $\sigma_+$  are filled
by a  vanishing fraction of states ($2(L-1)/(4L^2)$) and they belong to the
hidden spectrum \cite{jpb}, whereas $\sigma_0$ gives rise to the spectral
region of measure one, since the relevant  fraction of states is
$L(L-1)/L^2$. $\sigma_0$ reproduce the spectral density of a linear chain and
its states are completetely delocalized. On the other hand the states in 
$\sigma_-$ and $\sigma_+$, which characterize the typical behaviour of the comb
graph, are localized along the backbone presenting an exponential decay in the
direction of the finger.

The equation determining the fugacity $z$ for a finite comb is
\begin{equation}
\label{fill}
f= n_0(z,L) + n_{\sigma_-}(z,L)+ n_{\sigma_0}(z,L) +  n_{\sigma_+}(z,L)
\end{equation}
with $0\leq z\leq 1$.
Here $n_0(z,L)$, $n_{\sigma_+}(z,L)$ , $n_{\sigma_-}(z,L)$ and
$n_{\sigma_0}(z,L)$
represent respectively the fraction of particles in the ground state and
in the three spectral regions. More precisely
\begin{equation}
 n_{\sigma}(z,L)=\frac{1}{L^2}\sum_{E \in \sigma} d(E) \frac{z}{e^{\beta E}-z}
\end{equation}
where $d(E)$ denotes the degeneracy of the energy state $E$. Let us analyze
these four contributions separately.

In the thermodynamic limit the filling of the ground state,
\begin{equation}
n_0(z)=\lim_{L\to\infty}n_0(z,L)=\lim_{L\to\infty}\frac{1}{L^2}\frac{z}{1-z},
\label{n0}
\end{equation}
and the filling of the low hidden region,
\begin{equation}
n_{\sigma_-}(z)=\lim_{L\to\infty}n_{\sigma_-}(z,L)=\lim_{L \to \infty}
\frac{2}{L^2}\sum_{n=1}^{(L-1)/4}\left[
z^{-1}e^{\beta t(\sqrt{8} - 2\,\sqrt{1+\cos^2 \frac{2 \pi
n}{L}})}-1\right]^{-1} ,
\label{nsigma-}
\end{equation}
 are different from zero if and only if $\lim_{L\to \infty} z = 1$.
The filling of the spectral region of measure $1$ is given by:
\begin{equation}
n_{\sigma_0}(z)  = \lim_{L\to\infty}n_{\sigma_0}(z,L)=
\lim_{L\to\infty}\frac{1}{L}\sum_{n=1}^{L-1} \frac{z}{e^{\beta t(\sqrt{8}-
2\,\cos\frac{2 \pi n}{L})}-z} =  \frac{1}{2 \pi}\int_0^{2 \pi} dk
\frac{1}{z^{-1}e^{\beta t(\sqrt{8} - 2\,\cos k))}-1}=
\int { 1\over z^{-1}e^{\beta E}-1}\rho(E) dE
\label{nsigma0}
\end{equation}
$n_{\sigma_0}(z)$ is a finite positive number for each
value of $z$, $0\leq z\leq 1$.
For the high hidden region $\sigma_+$ we obtain:
\begin{equation}
0 \leq n_{\sigma_+}(z)=\lim_{L\to\infty}n_{\sigma_+}(z,L)
=\lim_{L\to \infty}\frac{2}{L^2}\sum_{n=1}^{(L-1)/4}
\frac{z}{e^{\beta t (\sqrt{8} + 2\,\sqrt{1+\cos^2 \frac{2 \pi n}{L}})}-z} <
\lim_{L\to\infty}\frac{2}{L} \frac{z}{e^{\beta t (\sqrt{8}+2)}-z} = 0
\label{nsigma+}
\end{equation}
Hence $n_{\sigma_+}(z,L)$ can be neglected.

Let us consider the behaviour of equation (\ref{fill}) in the thermodynamic
limit. Above the critical temperature it is satisfied by a value of $z=z'<1$
so that, according to equations (\ref{n0}) and (\ref{nsigma-}),
$n_0(z')=n_{\sigma_-}(z')=0$.
On the other hand for $T<T_C$ equation
(\ref{fill}) can be  solved only letting $z\to 1$ when $L\to\infty$.
In particular  we obtain a solution of equation (\ref{fill}) if and only if:
\begin{equation}
z \sim 1-\frac{\delta}{L^2} \ \ \ {\rm for\ \ }L\to\infty
\label{condz}
\end{equation}

Let us study in details the behaviour of $n_0(z)$ and $n_{\sigma_-}(z)$
when $z$ tends to $1$ as in equation (\ref{condz}). We get
\begin{equation}
n_{0}={1 \over \delta}
\label{n02}
\end{equation}
and the fraction of particles in $\sigma_-$ can be exactly evaluated by summing
the corresponding series:
\begin{equation}
n_{\sigma_-}=\lim_{L \to \infty} \frac{2}{L^2}
\sum_{n=1}^{(L-1)/4}\left[
e^{\beta t(\sqrt{8} - 2\,\sqrt{1+\cos^2 \frac{2 \pi
n}{L}})}-1+\frac{\delta}{L^2}\right]^{-1} =
\sum_{n=1}^{\infty} \frac{2}{2 \sqrt{2} \beta t\pi^2 n^2 +
\delta}= {{\rm csch}\, \alpha \, \left[ \alpha \cosh \alpha
-\sinh \alpha\right]\over \delta}
\label{nsigma-2}
\end{equation}
where
\begin{equation}
\alpha =
\frac{1}{2^{\frac{3}{4}}}\sqrt{\frac{\delta}{t \beta}}=
\frac{1}{2^{\frac{3}{4}}}\sqrt{\frac{T}{t n_0}}.
\label{alpha}
\end{equation}

Then below $T_C$ equation (\ref{fill}) for $z$  becomes an
equation for the new variable $\delta$,
\begin{equation}
\delta^{-1} +  \delta^{-1} \, {\rm csch}\, \alpha \, \left[ \alpha \cosh \alpha
-\sinh \alpha\right] = f- n_{\sigma_0}(1)=n_{TD}
\label{fill2}
\end{equation}
which can be numerically solved.
Hence, in order to obtain equation (\ref{fill2}) we had to evaluate the
thermodynamic limit of equations  (\ref{n0}),
(\ref{nsigma-}), (\ref{nsigma0}), (\ref{nsigma+}). We remark that only in the
case of  the spectral region of
measure one  it is possible to replace the sum  by an integral,  whereas in the
other cases the sums have to be   explicitly calculated.

Equation (\ref{fill2}) explains the main properties of BEC on the comb
lattice. First of all in this case $n_{TD}=f-n_{\sigma_0}(1)$  is the sum of two
contributions, the first due to the particles in the lowest energy state
$n_0=\delta^{-1}$, the other due to the particles in the low hidden
region, $n_{\sigma_-}=\delta^{-1} \, {\rm csch}\, \alpha \,
\left[ \alpha \cosh \alpha-\sinh \alpha\right]$.
Hence, on the comb graph, the filling of the ground state is smaller than
$n_{TD}$, as it has been numerically shown in figure 3.

A second point is that only states with arbitrary small energy
contribute to $n_{TD}$. Indeed the result of equation (\ref{nsigma-2}) does not
change if we force the index $n$ to be lower than $\epsilon L$, where
$\epsilon$ is an arbitrarily small fixed parameter. This means that,
$\forall\epsilon>0$, only states with energy smaller than
$t(\sqrt{8} - 2\,\sqrt{1+\cos^2 (2 \pi \epsilon)})$ contribute to $n_{TD}$.
Finally substituting $\delta^{-1}$ with $n_0$ in  equation (\ref{fill2}), we
obtain an exact relation between $n_0$ and $n_{TD}$. In figure 4 we checked
that the numerical results for the comb graph presented in the previous section
are perfectly reproduced by the exact calculation.
The dashed-dotted line is  obtained by first evaluating
$n_0$ and then using equation (\ref{fill2}) to get $n_{ TD}(n_0)$.
These data are very close both to $n_{\epsilon}$ (filling of a small energy
region around $E_0$) and to the theoretical value of $n_{TD}$.
\begin{figure}
\begin{center}
\epsfig{file=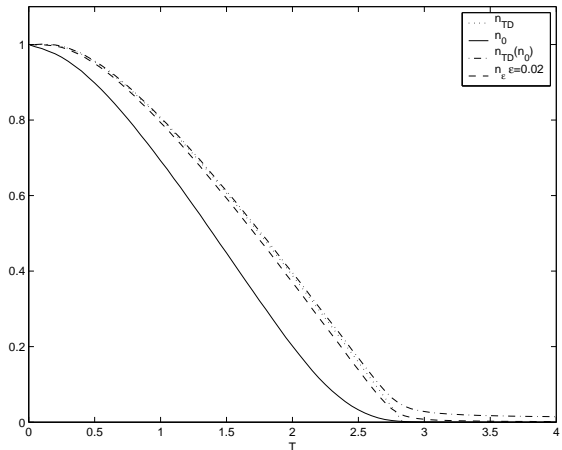, height=6cm,angle=0}
\caption{Check of equation (\ref{fill2}). The dotted line represents
the theoretical value of $n_{TD}$ obtained from (\ref{nTD}). The solid line is
a numerical evaluation of the filling of the ground state $n_0$. The dashed
line is a numerical evaluation $n_{\epsilon}$ for $\epsilon=0.02 t$. The
dash-dotted line is obtained evaluating $n_{TD}$ from (\ref{fill2}), where
$\delta$ and $\alpha$ are given by (\ref{alpha}), (\ref{n02}) and $n_0$ is
given by the numerical data of the solid line. The temperature is measured in
units of $t$ and numerical data are obtained from a comb of $N_r=40000$ sites.}
\end{center}
\end{figure}

\section {Coherence properties of the condensate}
The classical Bose-Einstein condensates on
homogeneous lattices present important large scale coherence
properties due to the macroscopic filling of a single quantum
state.  One of the most effective ways to put into evidence this
relevant feature of BEC is the study of the long distance
correlation function $C(i,j)$ defined by:
\begin{equation}
\label{correl}
C(i,j)=\sum_{k}{{\psi_k^r}^*(i)\psi_k^r(j)\over z^{-1}
e^{\beta E_k}-1}
\end{equation}
Where $\psi_k^r(j)$ is the eigenvector of $-tA^r_{ij}$ (\ref{hamsr}) of
eigenvalue $E_k^r$. In particular,  considering the correlation function
between two sites at a macroscopic distance
(i.e. $r_{ij}\sim r$, $r$ is the radius of the Van Hove sphere),
below the critical temperature in the thermodynamic limit we have:
\begin{equation}
C(i,j)\sim N_r n_{TD}{\psi_0^*}^r(i)\psi_0^r(j)\ \ \ {\rm for} \ \ r\to\infty
\label{corrsingle}
\end{equation}
Equation (\ref{corrsingle}) shows that below $T_C$ there are
$N_r  n_{TD}$ particles in the ground state $\psi_0^r(j)$. On regular lattices,
the wave function of the ground state is constant and the
correlation function at large distances does not depend on
$r_{ij}$. In particular for the three dimensional case ($N_r=L^3$) $C(i,j)\sim
(L^3 n_{TD})(L^3)^{-1}$; $(L^3 n_{TD})$ is the asymptotic number of particles of
the condensate and $(L^3)^{-1}$ is the normalizing factor given by the wave
function $\psi_0^r(j)$.

The large scale coherence properties of the Bose-Einstein condensate on a comb
graph can be studied by calculating the correlation functions between
the sites $i$ and $j$ of the backbone at a chemical distance $r_{ij}=d L$,
(this way if the size of the macroscopic system is
defined to be one, we are considering sites at distance $d$).
Let us evaluate the contribution to (\ref{correl}) of each single spectral
region by using the exact wave functions
$\psi_k^L(j)$ of the pure hopping model on a finite $L\times L$ periodic comb
\cite{jpb}, with $N_r=L^2$. For $E_0$ we have:
\begin{equation}
C_0(i,j)=C_0(d)={\psi_0^L}^*(i)\psi_0^L(j)\frac{z}{1-z}\sim
L^2n_0(\sqrt{2}L)^{-1}
\ \ \ {\rm for} \ \ L\to\infty
\label{corr0}
\end{equation}
$C_0(i,j)$ does not depend on the distance between the sites since the wave
function of the ground state is constant along the backbone. In (\ref{corr0})
$L^2n_0$ represents the filling of the state and $(\sqrt{2}L)^{-1}$ is the
normalizing factor of the wave function. In the low energy hidden spectrum we
have:
\begin{equation}
C_{\sigma_-}(d)=
\sum_{k=1}^{(L-1)/4} {\psi_k^L}^*(i)\psi_k^L(j) \left[
z^{-1}e^{\beta t(\sqrt{8} - 2\,\sqrt{1+\cos^2 \frac{2 \pi
k}{L}})}-1\right]^{-1} \sim {1\over(\sqrt{2}L)} \sum_{k=1}^{\infty}
{L^2 2 \cos(2\pi k d) \over \sqrt{2} \beta t\pi^2 k^2 +\delta}
\ \ \ {\rm for} \ \ L\to\infty
\label{corrsigma-}
\end{equation}
It should be noted that significant contribution to (\ref{corrsigma-}) arise
only from states with energy close to $E_0$, similar to what happens in equation
(\ref{nsigma-}).  Here the wave functions have
macroscopical oscillations along the backbone and the filling of each state
$L^2(\sqrt{2} \beta t\pi^2 k^2 +\delta)^{-1}$ is multiplied by
an oscillating factor $\cos(2\pi k d)$. $(\sqrt{2}L)^{-1}$ is  again the
normalizing factor of the wave functions.
In $\sigma_0$ and $\sigma_+$ we have $C_{\sigma_0}(i,j)=C_{\sigma_+}(i,j)=0$,
since in this case ${\psi_k^L}^*(i)\psi_k^L(j)$ is an oscillating factor with
diverging frequency giving rise to decoherence effects. The large scale
correlation function for the pure hopping model is then:
\begin{equation}
C(i,j)=C(d) \sim {L^2\over(\sqrt{2}L)}\left(n_0+ \sum_{k=1}^{\infty}
{2 \cos(2\pi k d) \over \sqrt{2} \beta t\pi^2 k^2 +\delta}\right)
={N_r\over(\sqrt{2 N_r})}\left(n_0+ \sum_{k=1}^{\infty}
{2 \cos(2\pi k d) \over \sqrt{2} \beta t\pi^2 k^2 +\delta}\right)
\ \ \ {\rm for} \ \ r=L\to\infty
\label{corrcomb}
\end{equation}
The two points correlation function diverges for $L\to\infty$, a
well known property of localized condensates.
Notice that each state of the condensate provides a different
oscillating contribution to (\ref{corrcomb}). In figure
5 we plot the correlation function $C(i,j)$ as a function of
the distance $d$.

\begin{figure}
\begin{center}
\epsfig{file=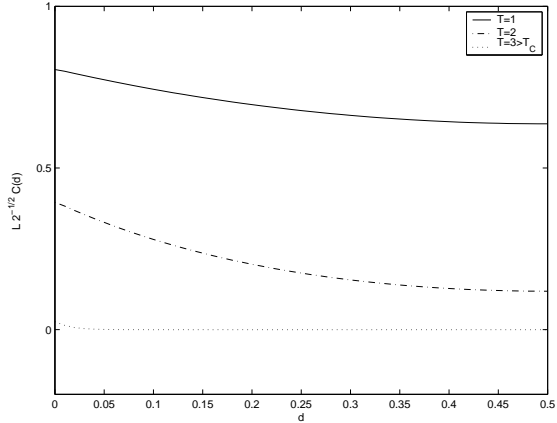, height=6cm,angle=0}
\caption{The correlation function given by (\ref{corrcomb}) as a function of the
macroscopic distance $d$. Above $T_C$, $C(d)$ is zero, while at low temperature
it depends on $d$.}
\end{center}
\end{figure}

We can also evaluate the correlation function between two sites of the same
finger. Since all the states of macroscopic filling
present the same behaviour along this direction (an exponential
decay depending on the distance from the backbone), in this case there are no
interference effects and one obtains an expression analogous to
(\ref{corrsingle}) ($|i|$ and $|j|$ represent the distance of the sites from the
backbone):
\begin{equation}
C(i,j)\sim {L^2\over(\sqrt{2}L)}\left(n_0 e^{-{\rm arcsh}(1)(|i|+|j|}+
\sum_{k=1}^{\infty}
{2 e^{-{\rm arcsh}(1)(|i|+|j|)} \over \sqrt{2} \beta t\pi^2 k^2 +\delta}\right)
={N_r n_{TD}\over(\sqrt{2 N_r})}e^{-(|i|+|j|){\rm arcsh}(1)}
\ \ \ {\rm for} \ \ r=L\to\infty
\label{corrcombfinger}
\end{equation}
$C(i,j)$ presents the diverging behaviour for $L\to\infty$ typical of localized
condensate.

\section{The general result}

As we have seen, in an inhomogeneous structure the occurrence of BEC does not
imply the macroscopic occupation of a single quantum state.
In this last section we prove the general condition  for a
bosonic hopping model on a graph to present condensation only on the lowest
energy state. In particular we will show that in a model presenting BEC at
finite temperature (i.e with a low energy hidden spectra or with a spectral
dimension greater than 2) one has condensation on the single state $E_0$ if the
spectrum $\rho(E)$ satisfies the following spectral condition:
\begin{equation}
\lim_{r\to\infty} \int {|{\rho'}^r(E)-\rho(E)| \over |E-E_0|} dE =0
\label{condsing}
\end{equation}
where
\begin{equation}
{\rho'}^r(E)={1\over{N_r}}{\sum_k}' \delta(E-E^r_k)
\label{rhoprime}
\end{equation}
The symbol ${\sum}'$ means that we are excluding from the sum the
contribution given by the ground state. Equation (\ref{condsing}) has a
clear mathematical interpretation. Indeed as (\ref{dens}) is the necessary
condition for evaluating the thermodynamic limit of the average value of any
bounded function $f(E_k)$ as $\int f(E) \rho(E) dE$, (\ref{condsing}) represents
the condition to evaluate by an integral even unbounded functions diverging in
$E_0$ as $(E-E_0)^{-1}$.  The general result can be obtained writing equation
(\ref{eq1}) as:
\begin{equation}
f={1 \over z^{-1}-1}+
\int_{E>\epsilon'} {{(\rho'}^r(E)-\rho(E))\over z^{-1}e^{\beta E}-1}dE +
\int_{E>\epsilon'} {\rho(E) \over z^{-1}e^{\beta E}-1}dE
\label{eq2}
\end{equation}
where we set $E_0=0$ and $\epsilon'$ is a generic energy between $E_0^r$
and $E_1^r$ the energy of the first excited state. Let us now take in
(\ref{eq2}) the thermodynamic limit $r\to\infty$.
For the first term $\lim_{r\to\infty}(z^{-1}-1)^{-1}=n_0$.
For the second term we have:
\begin{equation}
\int \left|{{(\rho'}^r(E)-\rho(E))\over e^{\beta E}-1}\right| dE\leq
\lim_{r\to\infty} \int {|{\rho'}^r(E)-\rho(E)| \over |E|} dE =0
\label{eq3}
\end{equation}
where we used the condition (\ref{condsing}). Finally in the last term
we let $\epsilon'\to 0$, obtaining:
\begin{equation}
n_0=f-
\int {\rho(E) dE\over z^{-1}e^{\beta E}-1}=n_{TD}
\label{eq4}
\end{equation}
and this implies that only the lowest energy state has a macroscopic filling.
The pure hopping model on the comb graph obviously does not satisfy 
(\ref{condsing}).
Notice that (\ref{condsing}) is not a condition on $\rho(E)$, the density of
states of the infinite structures, but a prescription on how $\rho^r(E)$ tends
to $\rho(E)$. The single state occupation then is determined by the behaviour of
the model on each large but finite mesoscopic scale.

\section{Concluding Remarks}
In this paper we address the issue of Bose-Einstein condensation
for a bosonic hopping model defined on the comb graph, a simple
inhomogeneous discrete structure. Such model is expected to
describe the low coupling limit for a comb-shaped network of
Josephson junctions (or, possibly, of optical traps confining
non-interacting atoms), a realistic system which is currently
being the subject of experimental research \cite{exp} . In
particular we show that the macroscopic occupation of the ground
state, which is one of the distinctive features of BEC on
(homogeneous) Euclidean lattices, is not sufficient for a complete
description of the phenomenon on the comb lattice. Indeed the
inhomogeneity of the network forces the macroscopic occupation of
a spectral region just above the ground state. This feature is
strictly related to the presence of the so-called {\it hidden
regions} in the spectrum of inhomogeneous structures such as the
comb graph, and, as we discuss in Section V, it is the cause of an
anomalous behaviour of the large distance two point correlation
functions. These results are obtained by exactly evaluating the
states filling on mesoscopic combs, where the energy
spectrum is nearly continuous, but the levels are still
distinguishable, being related to wave-functions which differ on a
large scale.
The conditions restoring the standard feature of the BEC, namely
the macroscopic occupation of a single quantum state, are
discussed as well.

\end{document}